%% file: main.tex
\documentclass[journal]{IEEEtran}
\input{myshorts}

\usepackage{cite,url}
\usepackage{amsmath,amssymb,amsfonts}
\usepackage{graphicx}
\usepackage{textcomp}
\usepackage{xcolor}
\usepackage{enumitem}
\usepackage{lipsum}
\usepackage{subcaption}
\usepackage{algorithm,algorithmic}
\renewcommand{\algorithmicrequire}{\textbf{Input:}}
\renewcommand{\algorithmicensure}{\textbf{Output:}}

 \newtheorem{definition}{Definition}
  
\begin{document}
\title{State Estimation  in Unobservable Power Systems via Graph Signal Processing Tools}

\author{\IEEEauthorblockN{Lital Dabush \IEEEmembership{Student~Member,~IEEE}, Ariel Kroizer, and Tirza Routtenberg \IEEEmembership{Senior~Member,~IEEE}}
\thanks{Lital Dabush, Ariel Kroizer, and Tirza Routtenberg are 
with the School of Electrical and Computer Engineering Ben-Gurion University of the Negev
 Beer-Sheva 84105, Israel.  e-mail: \{litaldab,arielkro\}@post.bgu.ac.il,~tirzar@bgu.ac.il. This research was supported by the Israel Ministry of National
Infrastructure, Energy, and Water Resources and by the BGU Cyber Security Research Center.}
\thanks{
This work has been submitted to the IEEE for possible publication.
Copyright may be transferred without notice, after which this version may
no longer be accessible.}
}

\maketitle
\begin{abstract}
We consider the problem of estimating the states in an unobservable power system. To this end, we propose novel graph signal processing (GSP) methods. 
For simplicity, we start with analyzing the DC power flow (DC-PF) model and then extend our algorithms to the AC power flow (AC-PF) model. The main assumption behind the proposed GSP approach is that the grid states, which include the vector of phases and the vector of the magnitudes of the voltages in the system, is a smooth graph signal with respect to the system admittance matrix that represents the underlying graph. Thus, the first step in this paper is to validate the graph-smoothness assumption of the states, both empirically and theoretically. Then, we develop the regularized GSP weighted least squares (GSP-WLS) state estimator, which does not require observability of the network. We propose a sensor  placement strategy that aims to optimize the estimation performance of the  GSP-WLS estimator. 
Finally, we extend the GSP-WLS estimator method to the AC-PF model by integrating a smoothness regularization term into the Gauss-Newton algorithm. 
Numerical results on the IEEE 118-bus system demonstrate that the new GSP methods outperform commonly-used estimation approaches 
and are robust to missing data.
\end{abstract}
\section{Introduction}
Power system state estimation (PSSE) is a critical component of modern Energy Management Systems (EMSs) for multiple  purposes, including monitoring,   analysis,   security, control, and   management of the power delivery \cite{Abur_Gomez_book}. 
The PSSE is conducted  using topological information, power measurements, and physical constraints \cite{Routtenberg_Tong_KCL} to estimate the voltages (states) at the system buses. 
The performance and reliability of the PSSE largely depend on the availability and the quality of the measurements \cite{Monticelli1999}.
However, there are various realistic scenarios where the system is {\em{unobservable}}, that is, 
 the number of sensors is not sufficiently
large or sensors are not well
placed in the network. In addition, a system may become unobservable
due to 
communication errors, topology changes,  sensor failures
 \cite{Abur_Gomez_book}, malicious attacks \cite{Kosut_Jia_Thomas_Tong_2011,drayer2018detection,morgenstern2020structuralconstrained}, and  electrical blackouts  \cite{buldyrev2010catastrophic}.
A direct implication of system unobservability is that estimators that assume deterministic states, such as the commonly-used weighted least-squares (WLS) estimator, can no longer be used since they are inaccurate,  inconsistent, and may have large estimation errors even in the absence of noise \cite{Gao2016Stefopoulos,Bayesian_tong}.
Therefore, 
developing new estimation 
methods that enable
the full functionality of  unobservable systems
is crucial for reliable operation of the power grid.

 State estimation in unobservable systems must incorporate
additional properties or information  beyond the  power flow equations in order to obtain a meaningful estimation.
Most existing approaches are
two-step solutions that first produce additional pseudo-measurements, e.g.
based on 
short-term load forecasting,
to make
the system observable, and then estimate the states  by any existing  technique
 \cite{5233865,6039370,Zhao2018La_Scala,Zhao2019Meliopoulos}.
 However, pseudo-measurements do not contain real-time data and thus, result in an inaccurate estimation.
Dynamic state estimation utilizes measurements  at different time instants \cite{Zhao2019Meliopoulos,Bhela2018Veeramachaneni}, but  need fast scan rates to capture the dynamics  and are based on restrictive stationary system assumptions  \cite{Zhao2019Meliopoulos}.
PSSE that uses data from  smart meters and phasor measurement units (PMUs)  to overcome observability issues \cite{Xu2004Abur,Gao2016Stefopoulos,zamzam2020modelfree} is usually inapplicable due to the limited deployment of these sensors \cite{deep_learning}, their  investment cost, and the  computational complexity of these approaches \cite{Zhang2011Huang}.
Sparse signal recovery methods \cite{Donti2020Zhang} use  matrix completion 
to estimate the states under low-observability conditions. However,
to employ the matrix completion framework,  the measurements matrix should  be a low-rank matrix. Unfortunately, this assumption is system-dependent and does not always hold due to, e.g. the spatial correlation between loads at neighboring buses.
Deep-learning techniques  have been recently used  for pseudo-measurement generation and for reconstructing missing  data for  PSSE \cite{deep_learning,zamzam2019physicsaware}. However,
the success of these techniques heavily depends on the availability of
numerous high-quality event labels that are rarely available in practice  \cite{Do_Coutto_Filho2007Schilling}.
 In addition, some of these methods  do not utilize the physical model \cite{deep_learning}, which may result in poor performance in physical systems.

In this paper, we consider the use of graph signal processing (GSP) tools  for state estimation in unobservable systems. 
A power system 
 can be represented as an undirected weighted graph, where the nodes (vertices) of the  graph
are the grid buses and its edges are the transmission lines.
In the literature, several  works have used concepts from graph theory
and graphical models in power systems for sensor placement \cite{Pourali2013Mosleh},
 topology identification \cite{Soltan2017Zussman,7541005,Grotas_Routtenberg_2019}, state estimation \cite{6687941,Giannakis_Wollenberg_2013,Kosut_Jia_Thomas_Tong_2011}, analysis
\cite{8347161,Guo_2018},
and optimal power flow calculation \cite{Dvijotham_2016,Caputo_2020}. However, the graphical model methods assume a specific statistical structure, which does not necessarily apply in power systems, and often do not use the physical models, which may result in poor performance in practice.
GSP is a new and emerging field that extends concepts and techniques from traditional digital signal processing (DSP) to data on graphs. GSP theory includes methods such as the Graph Fourier Transform (GFT), graph filters \cite{Sandryhaila_Moura_2013,8347162,Shuman_Ortega_2013}, and sampling and recovery of graph signals \cite{chen2015discrete,7352352}.
Recent works propose the integration of GSP in power systems, such as the GSP framework for the power grid  based on PMU data in
\cite{2021Anna}, the spectral graph analysis of power flows  in \cite{8832211}, and  false-data injection (FDI)
attacks detection by GSP in \cite{drayer2018detection}.
However,  state estimation without observability by GSP tools has not been demonstrated before.

In this paper, we develop GSP methods for state estimation in power systems based on interpreting  the voltage phases and magnitudes as graph signals.
First, 
 we show that the states  for static PSSE, that may include both the phases and the magnitudes of the voltages, are smooth graph signals with respect to (w.r.t.)  the nodal admittance matrix, which is a Laplacian matrix in the  graph representation of the network.
Second, we develop a  GSP-WLS estimation method for PSSE in the DC power flow (DC-PF) model that uses the smoothness of the states. The GSP-WLS estimator does not require observability of the network,  is only based on the current-time measurements, and does not assume a specific structure of the correlations between the buses. 
We also introduce a new approach for  sensor placement  that optimizes the estimation performance obtained by the GSP-WLS estimator.
It should be noted that
we initially treat a simplistic PSSE  in the DC-PF model in order to simplify  the presentation and the derivation of the estimation and sensor allocation methods.
Then, we extend our estimation method to the more realistic  AC-PF model by
developing a regularized Gauss-Newton method for PSSE that uses the smoothness of the phases and magnitudes of the voltages. 
Numerical simulations validate the merit of the new estimators under different setups.

The rest of this paper is organized as follows. In Section \ref{sec; Background} we introduce GSP background, the model, and the conventional estimation  approach.
In Section \ref{sub_sec_smooth} we study the GSP properties of the states. In Section \ref{sec;state_est}  we derive the  GSP-WLS state estimator for the DC-PF model and a sensor placement method that aims to optimize the  estimation performance. 
In Section \ref{sec;ac_model} we extend our estimation method to the AC-PF model by deriving the regularized Gauss-Newton method for state estimation. 
A simulation study is presented  in Section \ref{sec;Simulation} and  the conclusion appears in Section \ref{conclusion}.

In the rest of this paper, vectors and matrices are denoted by boldface lowercase letters and boldface uppercase letters, respectively. The notations $(\cdot)^T$, $(\cdot)^{-1
}$, $(\cdot)^{\dagger}$ and $\text{Tr}(\cdot)$ denote the transpose, inverse, Moore-Penrose pseudo-inverse, and trace operators,
respectively.
The $m$th element of the vector $\avec$  and the $(m, q)$th element of the matrix $\Amat$ are denoted by $a_m$ and $A_{m,q}$, respectively. Similarly, $\Amat_{\mathcal{S}_1,\mathcal{S}_2}$  denotes the submatrix of $\Amat$ whose rows and  columns are indexed by the sets $\mathcal{S}_1$ and  $\mathcal{S}_2$, where $\Amat_\mathcal{S}\define \Amat_{\mathcal{S},\mathcal{S}}$, and
$\avec_{\mathcal{S}}$ is a subvector of $\avec$ containing the elements indexed  by $\mathcal{S}$.
The gradient of a vector function $\gvec(\xvec)\in \mathbb{R}^{K} $ w.r.t. $\xvec\in\mathbb{R}^{M}$, $\frac{\partial\gvec(\xvec)}{\partial\xvec}$, is a matrix in $\mathbb{R}^{K\times M}$, with the $(k,m)$th element equal to $\frac{\partial g_k}{\partial x_m}$.
The matrices  $\Imat$ and
$\zerovec$ denote the identity matrix and
the  zero matrix with appropriate dimensions, respectively, and  $||\cdot||$ denotes the
Euclidean $l_2$-norm of vectors. For a vector $\avec$, ${\text{diag}}(\avec)$ is a diagonal matrix whose $n$th diagonal entry is $a_n$.

\section{Background and Model}
\label{sec; Background}
In this section, 
we first present the background on the theory of GSP in Subsection \ref{GSP_sec}.
Then, we present the considered power-flow 
measurement model, as well as the state estimation and network observability for this model,
in Subsection \ref{DCmodel_sec}.

\subsection{Background: GSP framework}
\label{GSP_sec}
Let ${\mathcal{G}}({\mathcal{V}},\xi)$ be a general undirected  weighted graph,
where ${\mathcal{V}}=\{1,\ldots,N\}$ and ${\xi}=\{1,\ldots,P\}$ are the sets of nodes and edges, respectively.
The matrix $\Wmat\in{\mathbb{R}}^{N\times N}$ is the weighted adjacency matrix of the graph $\mathcal{G}({\mathcal{V}},\xi)$, where $W_{k,n}$ denotes the weight of the edge between node $k$ and node $n$.
We assume  that $W_{k,n}\geq 0$ and that
$W_{k,n}= 0$ if no  edge exists between $k$ and $n$.
The graph Laplacian matrix is defined as 
\begin{equation}
\label{L_def}
\displaystyle L_{k,l} = \begin{cases}
	\displaystyle \sum\nolimits_{n=1}^N  W_{k,n},& k = l \\
  \displaystyle  -W_{k,l},& {\text{otherwise}} 
\end{cases},~~~k,l=1,\ldots,N.
\end{equation}
The Laplacian matrix, $\Lmat$, is a real and positive semidefinite matrix with the  eigenvalue-eigenvector decomposition
 \begin{equation}
\Lmat = \Vmat\Lambdamat \Vmat^T,   \label{SVD_new_eq}
 \end{equation}
where the columns of $\Vmat$ are the eigenvectors of $\Lmat$, $\Vmat^T=\Vmat^{-1}$, and $\Lambdamat \in {\mathbb{R}}^{N \times N}$ is a diagonal matrix consisting of the ordered eigenvalues of 
$\Lmat$: $0= \lambda_1 < \lambda_2 \leq \ldots \leq \lambda_N $. 
By analogy to the  frequency of signals in DSP,
 the Laplacian eigenvalues,
$\lambda_1,\ldots,\lambda_N$, can be interpreted as the graph frequencies, and the eigenvectors in
$\Vmat$  can be interpreted as the corresponding graph frequency components.
Together they define the spectrum of the graph ${\mathcal{G}}({\mathcal{V}},\xi)$ \cite{Shuman_Ortega_2013}.

A graph signal is a function that resides on a graph, assigning a scalar value to each node, and is 
an $N$-dimensional vector.
The GFT of a graph signal $\avec$  w.r.t. the graph ${\mathcal{G}}({\mathcal{V}},\xi)$ is defined as  \cite{Shuman_Ortega_2013,Sandryhaila_Moura_2013}:
 \begin{equation}
\label{GFT}
\tilde{\avec}\triangleq \Vmat^{-1}\avec.  
 \end{equation}
  Similarly,  the inverse GFT is obtained by left multiplication of a vector by $\Vmat$.
The Dirichlet energy of a graph signal, $\avec$, 
is defined as 
\beqna
{\mathcal{E}}_{\Lmat}(\avec)\define {\avec^T}\Lmat\avec &=&\frac{1}{2} \mathop \sum _{k=1}^N\sum_{n=1}^N W_{k,n}\big(a_k - a_n\big)^2
\label{eq:Dirichlet energy}\\
&=&\sum_{k=1}^N\lambda_k\tilde{a}_k^2,\label{eq:smoothness_def_GFT}
\eeqna where  the second equality is obtained by substituting  \eqref{L_def} and the last equality is obtained 
by substituting \eqref{SVD_new_eq} and \eqref{GFT}. 

The Dirichlet energy from \eqref{eq:Dirichlet energy} and \eqref{eq:smoothness_def_GFT} is a smoothness measure, which is used in graphs to quantify  the variability  encoded by the graph weights \cite{Shuman_Ortega_2013}.
A graph signal, $\avec$, is smooth if
 \begin{equation}
\label{smothness_full}
{\mathcal{E}}_{\Lmat}(\avec)\leq\varepsilon,
 \end{equation}
where $\varepsilon$ is small in terms of the specific application  \cite{Shuman_Ortega_2013}.
It can be seen that the smoothness condition in \eqref{smothness_full}  requires connected nodes to have similar values (according to \eqref{eq:Dirichlet energy}) and  forces the graph spectrum of the graph signal to be concentrated in the small eigenvalues region (according to \eqref{eq:smoothness_def_GFT}).

A graph filter applied on a graph signal is a linear operator   that satisfies the following \cite{8347162}:
\begin{equation}
\label{a_out_a_in}
   \avec_{out}=\Vmat {\text{diag}}(\psi(\lambda_1),\dots,\psi(\lambda_{N})) \Vmat^T\avec_{in},
\end{equation}
where $ \avec_{out}$ and $\avec_{in}$ are the output and input graph signals, and
  $\psi(\lambda_n)$ is the graph  filter  frequency response  at the graph frequency $\lambda_n$, $n=1,\ldots,N$.
Low-pass graph filters of order $K$ are defined as follows \cite{Rama2020Anna}.
\begin{definition}\label{def_LPF}
The graph filter 
 in \eqref{a_out_a_in}
is a  low-pass graph filter of order $K$
with a cutoff frequency at $\lambda_K$ 
if  $\eta_K<1$, where 
\begin{equation}
\label{eta_k_def}
    {\eta}_{k}\define \frac{\max\{\vert{\psi}({\lambda}_{k+1})\vert,\ldots,\vert{\psi}({\lambda}_{N})\vert\}}{\min\{\vert{\psi}({\lambda}_{1})\vert,\ldots,\vert{\psi}({\lambda}_{k})\vert\}},~ k=1,\ldots,N-1.
\end{equation}
\end{definition}
This definition implies that if $\eta_K<1$, then
 most of the energy of the graph filter is concentrated in the first $K$ frequency
bins of the graph filter \cite{Rama2020Anna}. 
Upon
passing a graph signal through $\Vmat {\text{diag}}(\psi(\lambda_1),\dots,\psi(\lambda_{N})) \Vmat^T$, the high-frequency
components (related to graph frequencies greater than $\lambda_K$) are attenuated relative to the low-frequency components (related to graph frequencies
lower than $\lambda_K$).
Thus, as long as the input of the filter is a ``well-behaved" excitation and does not possess strong high-pass components, the output signal is a $K$-low-pass graph signal \cite{Rama2020Anna}, and, thus, a {\em{smooth}} graph signal for small $K$, as defined in  \eqref{smothness_full}.



\subsection{DC-PF model: state  estimation and observability}
\label{DCmodel_sec}
A power system 
 is a network of buses (generators or loads) connected by transmission lines that can be represented as an undirected weighted graph, ${\mathcal{G}}({\mathcal{V}},\xi)$, where the set of nodes, $\mathcal{V}$, is the set of  $N$ buses  and the edge set, $\xi$, is the set of $P$ transmission lines between these buses.
We denote the set of all
sensor measurements by $\mathcal{M}$,  which includes
$M\define2P+N$ active power measurements at the $N$ buses and at the bi-directional $P$ transmission lines.
According to the  $\pi$-model \cite{Abur_Gomez_book},
each transmission line, $(k,n)\in \xi$, which connects buses $k$ and $n$, is characterized by an admittance value, $Y_{k,n}$.

The active power  and the voltages obey multivariate versions
of Kirchhoff’s and Ohm’s laws that result in the nonlinear  equations of the AC-PF model (see Section \ref{sec;ac_model}). In order to analyze the GSP properties and to simplify the presentation of the new  methods, we first approximate these equations by
the DC-PF model \cite{Abur_Gomez_book}, in which  the states are the voltage angles. 
Therefore, we consider first a  DC-PF  model with the following noisy  measurements of the active power \cite{Abur_Gomez_book}:
\begin{equation}
 \zvec =\Hmat\thetavec+\evec,\label{DC_model_vec}
\end{equation}
where
\begin{itemize}[leftmargin=0cm,labelsep=0.1cm,align=left]
    \item 
$\zvec = [z_1,\dots,z_M]^T \in {\mathbb{R}}^M$ is the active power vector.
\item  $\thetavec=[\theta_1,\dots,\theta_N]^T\in {\mathbb{R}}^N$ is the system state vector, 
where $\theta_n$ is the voltage angle at bus $n$.
In  low-observable systems, it is more convenient to delay the assignment of the reference angle (p. 165 in \cite{Monticelli1999}).
Thus,   $\thetavec$  includes the angle of the reference bus.
\item $\evec\in {\mathbb{R}}^M$  
is zero-mean Gaussian  noise
with  covariance  $\Rmat$.
\item $\Hmat\in {\mathbb{R}}^{M\times N}$ is the measurements matrix, which is determined by the topology of the network, the susceptance of the transmission lines, and the  meter  locations \cite{Kosut_Jia_Thomas_Tong_2011}. 
In particular,
 the $N$ rows of $\Hmat$ associated with the meters on the buses that measure  the total power flow on  transmission lines connected to these buses
create together the nodal admittance matrix (see, e.g. p. 97 in \cite{Monticelli1999}), $\Bmat$. $\Bmat$ has the following $(k,l)$-th element: 
\begin{equation}
\label{equ:YForm}
\displaystyle B_{k,l} = \begin{cases}
	\displaystyle \sum_{n\in{\mathcal{N}}_k } - b_{k,n},& k = l \\
  \displaystyle  b_{k,l},& (k, l) \in \xi \\
     \displaystyle 0,& \text{otherwise}
\end{cases},~\forall k,l=1,\ldots,N,
\end{equation}
where ${\mathcal{N}}_k$ is the set of buses  connected to bus $k$ and  $b_{k,n}<0$ is the susceptance of $(k,n)\in \xi$, i.e.  $b_{k,n}={\text{Im}}\{Y_{k,n}\}$.
\end{itemize}

The goal of  DC-PF PSSE is to recover the state vector, $\thetavec$, from the measurements vector, $\zvec$, for various monitoring purposes \cite{Abur_Gomez_book,Giannakis_Wollenberg_2013}.
Since  $\thetavec$ also includes the reference bus,
without loss of generality, we constrain the angle of bus $1$ (the reference bus) to be $\theta_1=0$.
The PSSE in this case is implemented by using the measurements described in \eqref{DC_model_vec} and the following WLS estimator \cite{Abur_Gomez_book}:
\beqna
\begin{array}{lr}
\hat{\thetavec}{}^{\text{WLS}}=\arg\min\limits_{\thetavecsmall\in \mathbb{R}^N}(\zvec-{\Hmat}{\thetavec})^T\Rmat^{-1}(\zvec-{\Hmat}{\thetavec})
\\
{\text{ such that }}\theta_1=0.
\end{array} 
\label{WLS}
\eeqna
The solution of \eqref{WLS} is
\beqna
\label{WLS2}
\left\{\begin{array}{l}\hat{\thetavec}{}_{\bar{\mathcal{V}}}^{\text{WLS}}=\Kmat \zvec\\
\hat{\theta}_1^{\text{WLS}}=0\end{array}\right.,
\eeqna
where 
\be
\label{K_def}
\Kmat\define (\Hmat_{\mathcal{M},\bar{\mathcal{V}}}^T\Rmat^{-1}\Hmat_{\mathcal{M},\bar{\mathcal{V}}})^{-1}\Hmat_{\mathcal{M},\bar{\mathcal{V}}}^T\Rmat^{-1}
\ee 
and $\bar{\mathcal{V}}\define \mathcal{V}\setminus1$ is the set of all
buses except the reference bus.

 For state estimation to be feasible, we need
to have enough  measurements such that the system state
can be uniquely determined by the WLS estimation approach.
 This observability requirement, before  the
assignment of  reference angles, can be defined by
 one of the following  
 (p. 165 in \cite{Monticelli1999}).
 \begin{definition}
 \label{def_obs}
 Assume the DC-PF model from \eqref{DC_model_vec}. The network is observable
if any matrix that is obtained from $\Hmat$ by deleting one of its columns has a full column rank of $N-1$.  
 \end{definition}
  \begin{definition}
 \label{def_obs2}
The network is observable if the following holds: $\Hmat\thetavec=\zerovec$ 
{\em{if and only if}} $\thetavec=\alpha {\mathbf{1}}$, where $\alpha$ is an arbitrary scalar.
\end{definition}
In particular, since  according to Definition \ref{def_obs} 
$\Hmat_{\mathcal{M},\bar{\mathcal{V}}}$  has full
column rank for an observable system, observability ensures that $\Hmat_{\mathcal{M},\bar{\mathcal{V}}}^T\Rmat^{-1}\Hmat_{\mathcal{M},\bar{\mathcal{V}}}$  is  non-singular, and the WLS estimator from \eqref{WLS2}-\eqref{K_def}  is well-defined for any observable network.

In practice, however, network observability is not always guaranteed. In such cases, the WLS estimator from \eqref{WLS2} cannot be implemented.
Even for observable systems,
errors and  outliers
may have a disastrous effect on the state estimation. In the following, we show that incorporating graphical information by using GSP tools improves  the state estimation performance and enables estimation even  in unobservable systems.

\section{GSP properties of the states}\label{sub_sec_smooth}
The power system can be  represented  as an undirected weighted graph, ${\mathcal{G}}({\mathcal{V}},\xi)$, as described at the beginning of Subsection \ref{DCmodel_sec}.
In this context, the state vector, $\thetavec\in{\mathbb{R}}^N$,  and the subvector of $\zvec$ from \eqref{DC_model_vec} that contains the $N$ active power injection measurements at the $N$ buses, denoted as $\zvec_{\text{bus}}$, can be  interpreted as  {\em{graph signals}}.
 In this graph representation, 
the nodal admittance matrix from \eqref{equ:YForm} is a Laplacian matrix:
\be
\label{L_is_B}
\Lmat=\Bmat.
\ee

In this section, we  establish the  graph low-pass nature 
 and the smoothness of 
 the state vector in power systems under normal operation conditions and where the Laplacian matrix is set to be the nodal admittance matrix, as defined in \eqref{L_is_B}.
That is, by using the smoothness defined in \eqref{eq:Dirichlet energy}-\eqref{smothness_full}, we show that 
 \begin{equation}
\label{smothness_full_theta}
{\mathcal{E}}_{\Lmat}(\thetavec)=
{\thetavec^T}\Lmat\thetavec\leq\varepsilon,
 \end{equation}
where $\varepsilon$ is small relative to the other parameters in the system.
These results are consistent with the low-pass graph nature of the complex voltages in the AC-PF model, described in \cite{2021Anna}.
\subsection{Theoretical validation - Output of a low-pass graph filter}
First, we show analytically that the state vector is a low-pass graph signal. 
By substituting \eqref{L_is_B} in the model in \eqref{DC_model_vec}, after taking only the active power injection measurements we obtain 
\beqna\label{eq;dc_model_injections}
    \zvec_{\text{bus}}=\Lmat\thetavec+\evec_{\text{bus}},
\eeqna
where $\evec_{\text{bus}}$ contains the elements of the noise vector, $\evec$, that are related to the $N$  power measurements at the $N$ buses. Equation \eqref{eq;dc_model_injections} implies that 
since $\Lmat$ is a Laplacian matrix, it satisfies $\Lmat\onevec=\zerovec$ \cite{Newman_2010},
where $\onevec$ is the a vector of ones with appropriate dimensions. Thus, we can recover the states from \eqref{eq;dc_model_injections} up-to-a-constant shift, which can be written as
\beqna
\label{dc_z_bus}
    \thetavec&=&\Lmat^{\dagger}\left(\zvec_{\text{bus}}-\evec_{\text{bus}}\right)+c_1\onevec
    \nonumber\\&=&
    \Vmat\Lambdamat^{\dagger} \Vmat^T\left(\zvec_{\text{bus}}-\evec_{\text{bus}}\right)+c_1\onevec,
\eeqna
where $c_1$ is an arbitrary constant that represents the constant-invariant property of the state vector \cite{Abur_Gomez_book,Kroizer2019Routtenberg}
and the last equation is obtained by substituting \eqref{SVD_new_eq}.
Without loss of generality, we choose the value of  $c_1$  to be
\begin{equation}
    c_1=c_2\onevec^T\left(\zvec_{\text{bus}}-\evec_{\text{bus}}\right),
    \label{c2}
\end{equation}
where $c_2$ is an arbitrary constant. Using \eqref{c2} and  the definition of the pseudo-inverse operator,  the model in \eqref{dc_z_bus} can be written as the following linear graph-filter  input-output model:
\begin{equation}
\label{dc_z_bus_invertable}
    \thetavec=\Vmat {\text{diag}}(\psi(\lambda_1),\dots,\psi(\lambda_{N}))\Vmat^T \left(\zvec_{\text{bus}}-\evec_{\text{bus}}\right),
\end{equation}
where
\be\label{freq_res}
\psi(\lambda_n)=\left\{\begin{array}{lr}
\frac{1}{\lambda_n},&n=2,\ldots,N\\
Nc_2,&n=1
\end{array}\right..
\ee
That is, $\thetavec$ is an output of a  graph filter with the  graph frequency response in \eqref{freq_res}.
This graph-filter representation holds under the assumption that the network is  connected. Thus, $\lambda_1$ is the only zero eigenvalue of $\Lmat$  with the associated eigenvector $\frac{1}{\sqrt{N}}\onevec$ \cite{Newman_2010}.


Since the eigenvalues of $\Lmat$ are ordered,  $0=\lambda_1<\lambda_2\leq\lambda_3\leq \ldots\leq\lambda_N$,
it can be seen that the graph frequency response in \eqref{freq_res} decreases as $n$ increases, as long as $c_2>\frac{1}{N\lambda_2}$.
By substituting \eqref{freq_res} in \eqref{eta_k_def}, we obtain
\be\label{eta_k}
{\eta}_{k}=\left\{\begin{array}{lr}
\frac{\lambda_{k}}{{\lambda}_{k+1}},&2 \leq k \leq N-1\\\frac{1}{Nc_2\lambda_2},&k=1
\end{array}\right.,
\ee
where $\eta_k<1$, $k=2,\ldots,N-1$.
By choosing $c_2>>\frac{1}{N\lambda_2}$,
we obtain that  ${\eta}_{1}<<1$.
Thus, according to Definition \ref{def_LPF}, $\Vmat {\text{diag}}(\psi(\lambda_1),\dots,\psi(\lambda_{N})) \Vmat^T$ in \eqref{dc_z_bus_invertable} is a graph low-pass filter of any order $K\geq 1$.
 Since $\zvec_{\text{bus}}$ in \eqref{dc_z_bus} includes the generated powers and loads, it can be assumed to be random \cite{Aien2012Aminifar,Chakhchoukh_2011}, and thus,
 the input signal, $\zvec_{\text{bus}}-\evec_{\text{bus}}$,  does not possess strong high-pass components.  Thus, as explained after Definition \ref{def_LPF}, the state vector $\thetavec$
is a first-order low-pass graph signal, and   a smooth graph signal, as defined in \eqref{smothness_full_theta}.

\subsection{Experimental validation in IEEE  systems}
\begin{figure}[htb]
		\begin{center}
		\centering\includegraphics[width=4.5cm]{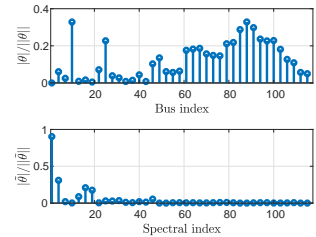}
  \caption{The state vector (top) and its GFT (bottom) for IEEE 118-bus system.}
		\label{fig30}	
		\end{center}
	\end{figure}

\begin{figure}[htb]
		\begin{center}
		\centering\includegraphics[width=4.5cm]{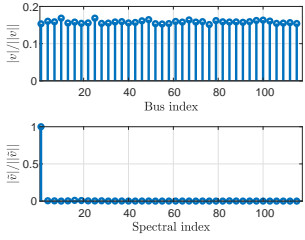}
  \caption{The voltage magnitude vector (top) and its GFT (bottom) for IEEE 118-bus system.}
		\label{fig:gft_v_118}	
		\end{center}
	\end{figure}

\begin{figure}[htb]
		\begin{center}
		\centering\includegraphics[width=4.5cm]{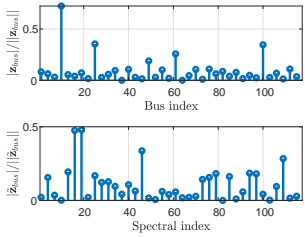}
  \caption{The active power measurements vector (top) and its GFT (bottom) for IEEE 118-bus system.}
		\label{fig31}	
		\end{center}
	\end{figure}

In the following, we demonstrate the smoothness of the state signal 
in the graph spectrum domain 
for  the IEEE test case systems \cite{iEEEdata}. We also demonstrate the smoothness of the voltage magnitude vector, $\vvec\define[|v_1| ,\dots,|v_N|]^T$, which can be interpreted  as  {\em{graph signals}}, that  will be used in Section \ref{sec;ac_model} along with the AC-PF model.
In Fig. \ref{fig30}
we compare between 
the normalized  state vector, $\frac{\thetavecsmall}{||\thetavecsmall||}$, and its GFT (calculated   by using \eqref{GFT}),
$\frac{\tilde{\thetavecsmall}}{||\tilde{\thetavecsmall}||}$,   versus bus or spectral indices, for the IEEE 118-bus system \cite{iEEEdata}.
Similarly, in Fig. \ref{fig:gft_v_118}  we present
the normalized voltage magnitude vector, $\frac{\vvec}{||\vvec||}$, and its GFT, $\frac{\tilde{\vvec}}{||\tilde{\vvec}||}$, and
in Fig. \ref{fig31} we present
the normalized  power vector, $\frac{\zvec_{\text{bus}}}{||\zvec_{\text{bus}}||}$, and its GFT, $\frac{\tilde{\zvec}_{\text{bus}}}{||\tilde{\zvec}_{\text{bus}}||}$.
For the sake of clarity,
the vectors in Figs.  \ref{fig30}-\ref{fig31} have been decimated by a factor of $3$. 

	It can be seen that most of the energy of the state signal, i.e. the phases  (Fig. \ref{fig30}) and the magnitudes  (Fig. \ref{fig:gft_v_118}) of the voltages, is concentrated in the low graph frequencies region. Thus, we can conclude that the state vector and the voltage magnitude vector are smooth graph signals  in the sense of   \eqref{eq:smoothness_def_GFT}. In contrast, the energy of the power injection measurements vector (Fig. \ref{fig31}) is uniformly distributed across  all  graph frequencies. Thus, the power signal cannot considered to be smooth. Similar results were obtained for other IEEE systems. 

Next, we validate experimentally that the states, $\thetavec$, and the magnitudes, $\vvec$, are significantly smoother than the power vector, $\zvec_{\text{bus}}$, by comparing their associated normalized Dirichlet energy 
for typical
 IEEE systems in Table \ref{tabSmooth}.  The values of the  Laplacian (nodal admittance) matrix, $\Lmat=\Bmat$, the voltages, and the power data are  taken from   \cite{iEEEdata}.
 It can be seen that the phases and magnitudes are much smoother than the power injection  vectors. This result is reasonable since the phase differences between connected buses are   small under normal conditions and the magnitudes are approximately constant \cite{Giannakis_Wollenberg_2013}, while the power may be very different since each generator/load injects  different power into the system.

\begin{table}[hbt]
\begin{center}
\begin{tabular}{|c|c|c|c|c|c|}
\hline
\text{}&\multicolumn{5}{|c|}{\textbf{IEEE test-case system}} \\
\cline{2-6} 
Measure & \textbf{\textit{14-bus}}& \textbf{\textit{30-bus}}& \textbf{\textit{57-bus}}& \textbf{\textit{118-bus}}& \textbf{\textit{300-bus}} \\
\hline
$\frac{{\mathcal{E}}_{\Lmat}(\thetavecsmall)}{||\thetavecsmall||^2}$ &0.6617 &0.3015 &0.3714 &1.1740 & 1.2371\\
\hline
$\frac{{\mathcal{E}}_{\Lmat}(\vvec)}{||\vvec||^2}$ &0.0036 &0.0022 &0.008 &0.0082 & 0.0199\\
\hline
$\frac{{\mathcal{E}}_{\Lmat}(\zvec_{\text{bus}})}{||\zvec_{\text{bus}}||^2}$ &16.4079 & 18.3307 & 50.8035 & 56.1047 & 138.8024 \\
\hline
\end{tabular}
\caption{Normalized smoothness values of IEEE  systems.} 
\label{tabSmooth}
\end{center}
\end{table}

\section{GSP-WLS estimator in DC-PF model}
\label{sec;state_est}
The recovery of smooth graph signals by incorporating regularization terms has been well studied in the GSP literature  \cite{reconstruction_of_smooth_graph_signals,8347162} and in the context of Laplacian regularization \cite{laplacianregularization,laplacianreg}.
In this section, we cast the state estimation problem  as a regularized graph
signal recovery problem.
In particular, we exploit the smoothness of the state vector, established in Section \ref{sub_sec_smooth},
to  develop  the smoothness-based regularized GSP-WLS estimator of the states in Subsection \ref{subsec;PSSE}. We discuss the properties of the proposed approach in Subsection \ref{subsec;remarks}, where its
 main advantage is that it
 does not require system observability.
 In Subsection \ref{power} we introduce a by-product of this approach: an estimator of the missing power data.
 Finally, in Subsection \ref{sec;sampling}
 we design a sensor allocation policy
 that aims to optimize the performance  of the GSP-WLS estimator.
\subsection{GSP-WLS estimator for partial measurement model}\label{subsec;PSSE}
In the following, 
we consider the case where only partial observations of the signal  $\zvec$ from \eqref{DC_model_vec} are available
over  a subset of sensors from ${\mathcal{M}}$, where this subset is denoted by $\mathcal{S}$ and $\mathcal{S}\subseteq {\mathcal{M}}$. A sensor at a particular location
provides one row in the measurements matrix, $\Hmat$. Thus, based on the  model in \eqref{DC_model_vec}, the partial measurement vector can be written  as
\begin{equation}
\zvec_{\mathcal{S}}=\Hmat_{\mathcal{S},\mathcal{V}}{\thetavec}+\evec_{\mathcal{S}}.\label{reduced model}
\end{equation}
Since $\evec_{\mathcal{S}}$ contains  the elements of the noise vector, $\evec$, of the set of available measurements, $\mathcal{S}$, it 
is a zero-mean Gaussian noise vector
 with a covariance matrix $\Rmat_{\mathcal{S}}$.
If the columns of $\Hmat_{\mathcal{S},\mathcal{V}}$ after deleting one  column are linearly dependent, then,  from Definition \ref{def_obs},
 the new model 
in \eqref{reduced model} with the matrix $\Hmat_{\mathcal{S},\mathcal{V}}$ is {\em{unobservable}}. 
In this case, the WLS estimator for the model in \eqref{reduced model}
cannot be developed
 similarly to in \eqref{WLS}-\eqref{WLS2}, since,
according to Definition \ref{def_obs2}, the state, $\thetavec$,
cannot be uniquely (up-to-a-constant) determined 
from \eqref{reduced model}.

As a result, we need to incorporate
additional properties  beyond the  power flow equations in \eqref{reduced model} to obtain a valid state estimation.
Here, we propose to recover $\thetavec$ by using the GSP-WLS estimator that incorporates the smoothness constraint from \eqref{smothness_full_theta}. Thus, the GSP-WLS estimator is defined by
\beqna\label{eq:Smooth_wls}
\hspace{-0.2cm}\hat{\thetavec}{}^{\text{GSP-WLS}}\hspace{-0.1cm}=\hspace{-0.1cm}\arg\min_{{{\thetavecsmall}\in \mathbb{R}^{N}}} (\zvec_{\mathcal{S}}-\Hmat_{\mathcal{S},\mathcal{V}}{\thetavec})^T\Rmat_{\mathcal{S}}^{-1}(\zvec_{\mathcal{S}}-\Hmat_{\mathcal{S},\mathcal{V}}{\thetavec})\nonumber\\
{\text{such that }} 1)~\theta_1=0 {\text{ and }} 2)~{\thetavec}^T {\Lmat}  {\thetavec} \leq \varepsilon.
\eeqna
By using $\bar{\mathcal{V}}= \mathcal{V}\setminus1$,  ${\thetavec}_{\bar{\mathcal{V}}}$ is the state vector without the  reference bus state, $\theta_1$,
and
$\Lmat_{\bar{\mathcal{V}}}$ is the submatrix of $\Lmat$ obtained by removing its first row and   column.
The smoothness constraint in \eqref{smothness_full_theta} after substituting $\theta_1=0$ can be rewritten as \beqna\label{eq:smooth_reduce}{\thetavec}_{\bar{\mathcal{V}}}^T\Lmat_{\bar{\mathcal{V}}}{\thetavec}_{\bar{\mathcal{V}}}\leq\varepsilon.
\eeqna 
By using the smoothness constraint from \eqref{eq:smooth_reduce} and substituting $\theta_1=0$
in \eqref{eq:Smooth_wls},
 the GSP-WLS estimator is given by
\beqna\label{eq:smooth_wls_reduced}
\hat{\thetavec}{}_{\bar{\mathcal{V}}}^{\text{GSP-WLS}}=\hspace{5.75cm}\nonumber\\\arg\min_{{{\thetavecsmall}_{\bar{\mathcal{V}}}\in \mathbb{R}^{N-1}}} (\zvec_{\mathcal{S}}-\Hmat_{\mathcal{S},\bar{\mathcal{V}}}{\thetavec}_{\bar{\mathcal{V}}})^T\Rmat_{\mathcal{S}}^{-1}(\zvec_{\mathcal{S}}-\Hmat_{\mathcal{S},\bar{\mathcal{V}}}{\thetavec}_{\bar{\mathcal{V}}})\nonumber\\
{\text{such that }} {\thetavec}_{\bar{\mathcal{V}}}^T {\Lmat}_{\bar{\mathcal{V}}}  {\thetavec}_{\bar{\mathcal{V}}} \leq \varepsilon,\hspace{4cm}
\eeqna and $\hat{\theta}{}_1^{\text{GSP-WLS}}=0$. 
Then,
by using the Karush-Kuhn-Tucker (KKT) conditions \cite{Boyd_2004}, the minimization problem in \eqref{eq:smooth_wls_reduced} can be replaced by the following regularized optimization problem 
(see, e.g. pp. 17-19 in \cite{vanwieringen2020lecture}):
\beqna\label{eq:lagrange_pro}
\hat{\thetavec}{}_{\bar{\mathcal{V}}}^{\text{GSP-WLS}}=\arg\min_{{{\thetavecsmall}_{\bar{\mathcal{V}}}\in \mathbb{R}^{N-1}}} \{(\zvec_{\mathcal{S}}-\Hmat_{\mathcal{S},\bar{\mathcal{V}}}{\thetavec}_{\bar{\mathcal{V}}})^T\Rmat_{\mathcal{S}}^{-1}\hspace{0.25cm}\nonumber\\\times(\zvec_{\mathcal{S}}-\Hmat_{\mathcal{S},\bar{\mathcal{V}}}{\thetavec}_{\bar{\mathcal{V}}})+\mu{\thetavec}_{\bar{\mathcal{V}}}^T {\Lmat}_{\bar{\mathcal{V}}}  {\thetavec}_{\bar{\mathcal{V}}}\},
\eeqna
and
$\hat{\theta}{}_1^{\text{GSP-WLS}}=0$. The term ${\thetavec}_{\bar{\mathcal{V}}}^T {\Lmat}_{\bar{\mathcal{V}}}  {\thetavec}_{\bar{\mathcal{V}}}$ is a regularization term, which is based on the smoothness constraint from \eqref{eq:smooth_reduce}.
The parameter $\mu\geq 0$ is a Lagrange multiplier,  which is a tuning parameter that replaces $\varepsilon$ as a regularization parameter and is discussed in  Subsection \ref{subsec;remarks}.
If the system is unobservable based on the sensors at $\mathcal{S}$, then  we have to choose $\mu>0$.
The GSP-WLS estimator that solves \eqref{eq:lagrange_pro} is obtained by equating the derivative of \eqref{eq:lagrange_pro} w.r.t. ${\thetavec}_{\bar{\mathcal{V}}}$ to zero 
which results in \cite{vanwieringen2020lecture}
\beqna
\label{estimator}
\left\{\begin{array}{l}\hat{\thetavec}{}_{\bar{\mathcal{V}}}^{\text{GSP-WLS}}=\tilde{\Kmat}(\mathcal{S},\mu)\zvec_{\mathcal{S}}\\
\hat{\theta}{}_1^{\text{GSP-WLS}}=0\end{array}\right.,
\eeqna
where 
\be
\label{estimator_matrix}
\tilde{\Kmat}(\mathcal{S},\mu)
\define(\Hmat_{\mathcal{S},\bar{\mathcal{V}}}^T\Rmat_{\mathcal{S}}^{-1}\Hmat_{\mathcal{S},\bar{\mathcal{V}}}+\mu\Lmat_{\bar{\mathcal{V}}})^{-1}\Hmat_{\mathcal{S},\bar{\mathcal{V}}}^T\Rmat_{\mathcal{S}}^{-1}.
\ee
The proposed estimator in \eqref{estimator}-\eqref{estimator_matrix} is named the GSP-WLS estimator in the following.
For an unobservable system, the matrix $\Hmat_{\mathcal{S},\bar{\mathcal{V}}}^T\Rmat_{\mathcal{S}}^{-1}\Hmat_{\mathcal{S},\bar{\mathcal{V}}}$ is a singular matrix and the additional  term in \eqref{estimator_matrix},
$\mu\Lmat_{\bar{\mathcal{V}}}$ with $\mu> 0$, enables the matrix inversion and improves the numerical stability of the GSP-WLS estimator. 
\subsection{Discussion}
\label{subsec;remarks}
The main advantage of the proposed GSP-WLS estimator in \eqref{estimator}-\eqref{estimator_matrix} is that it
 does not require observability of the system.
 This estimator is a function of the regularization parameter, $\mu \geq 0$,
 that can be determined {\em{a-priori}} according to historical values of $\varepsilon$ in \eqref{eq:Smooth_wls} 
 or  by trial and error.
In the following, we present special cases of the proposed GSP-WLS estimator.

\begin{enumerate}[leftmargin=0.35cm,labelsep=0.5cm,align=left]
    \item\label{special_case1} \underline{Full observability:} 
If we have access to all sensors, then, by substituting 
$\mathcal{S}=\mathcal{M}$ and
 $\mu=0$ in
\eqref{estimator_matrix} we obtain that
\be
\tilde{\Kmat}(\mathcal{M},0)=\Kmat,
\ee where $\Kmat$
is defined in \eqref{K_def}.
Thus,  if $\mathcal{S}=\mathcal{M}$, then 
 the GSP-WLS estimator from \eqref{estimator} with $\mu=0$ coincides with the  WLS estimator, $\hat{\thetavec}{}^{\text{WLS}}$, from \eqref{WLS2}.
\item \underline{Large $\mu$:} 
At the other extreme,  for $\mu\rightarrow \infty$, the coefficient matrix from \eqref{estimator_matrix} satisfies
$
\lim_{\mu\to \infty}\tilde{\Kmat}(\mathcal{S},\mu)
=\zerovec$.
Thus, in this case, the GSP-WLS estimator from \eqref{estimator} satisfies 
$\hat{{\thetavec}}{}^{\text{GSP-WLS}}\rightarrow\zerovec$. This zero estimator can be interpreted as the {\em{a-priori}}  state estimator, which does not use the observations. Thus, taking too large value of $\mu$ is unhelpful.
\item \underline{Relation with the pseudo-measurement WLS (pm-WLS)} {\underline{estimator:}}
The pm-WLS estimator for unobservable systems is based on 
generating  pseudo-measurements  of typical power injection/consumption values from  historical data  \cite{Abur_Gomez_book},  \cite{Do_Coutto_Filho2007Schilling}.
In this case,  the received measurements are processed together with {\em{a-priori}} estimated (predicted) states (without the reference bus),  $\hat{\thetavec}_{prior}\in\mathbb{R}^{ N-1}$, which are assumed to have  the forecasting error covariance matrix, $\Rmat_{prior}\in\mathbb{R}^{ (N-1)\times (N-1)}$. 
The pm-WLS estimation can be seen as the  {\em{a-posteriori}} estimated  state \cite{5233865}:
\beqna
\label{eq:pseudo_estimator}
\left\{\begin{array}{l}
\hat{\thetavec}^{({\text{pm-WLS}})}_{\bar{\mathcal{V}}}=\Kmat_1 \zvec_{\mathcal{S}}+
\Kmat_2\hat{\thetavec}_{prior}\\
\hat{\theta}_1=0\end{array}\right.,
\eeqna
where 
\beqna\label{K_1}
\Kmat_1&=&({\Hmat}_{\mathcal{S},\bar{\mathcal{V}}}^T\Rmat_{\mathcal{S}}^{-1}{\Hmat}_{\mathcal{S},\bar{\mathcal{V}}}+\Rmat_{prior}^{-1})^{-1}{\Hmat}_{\mathcal{S},\bar{\mathcal{V}}}^T\Rmat_{\mathcal{S}}^{-1}
\\
\label{K_2}
\Kmat_2&=&({\Hmat}_{\mathcal{S},\bar{\mathcal{V}}}^T\Rmat_{\mathcal{S}}^{-1}{\Hmat}_{\mathcal{S},\bar{\mathcal{V}}}+\Rmat_{prior}^{-1})^{-1}\Rmat_{prior}^{-1}.
\eeqna
It can be seen that if  $\hat{\thetavec}_{prior}\hspace{-0.15cm}=\hspace{-0.15cm}\zerovec$  and $\Rmat_{prior}^{-1}\hspace{-0.15cm}=\hspace{-0.15cm}\mu\Lmat_{\bar{\mathcal{V}}}$, then $\Kmat_1=\tilde{\Kmat}(\mathcal{S},\mu)$
and the pm-WLS estimator in \eqref{eq:pseudo_estimator} coincides with the GSP-WLS estimator in \eqref{estimator}. 
Thus, the proposed GSP-WLS estimator can be interpreted as a special case of the pm-WLS estimator, where the GSP theory gives a mathematical way to determine the pseudo-data information.
\end{enumerate}

\subsection{Estimation of missing power  measurements}
\label{power}
An important straightforward by-product of the GSP-WLS estimator is the following method for reconstructing the missing data
 of active power measurements. 
In the unobservable system, we have measurements obtained from  the set of sensors, $\mathcal{S}$, which is given by $\zvec_{\mathcal{S}}$. Our goal in this subsection is to recover the other measurements that are included in the vector 
$\zvec_{{\mathcal{M}}\setminus\mathcal{S}}$.
Based on the  model in \eqref{DC_model_vec}, similar to \eqref{reduced model}, the unobservable measurement vector can be written  as
\begin{equation}
\zvec_{{\mathcal{M}}\setminus\mathcal{S}}=\Hmat_{{{\mathcal{M}}\setminus\mathcal{S}},\mathcal{V}}{\thetavec}+\evec_{{{\mathcal{M}}\setminus\mathcal{S}}},\label{reduced model2}
\end{equation}
 where $\evec_{{{\mathcal{M}}\setminus\mathcal{S}}}$ 
is a zero-mean noise vector
 with a covariance matrix $\Rmat_{{{\mathcal{M}}\setminus\mathcal{S}}}$.
 By substituting  the GSP-WLS estimator from \eqref{estimator} in \eqref{reduced model2} and removing the noise term, we obtain 
 the following WLS-type estimator of the missing power measurements:
 \begin{equation}
\hat{\zvec}_{{\mathcal{M}}\setminus\mathcal{S}}=\Hmat_{{{\mathcal{M}}\setminus\mathcal{S}},\mathcal{V}}\hat{\thetavec}{}^{\text{GSP-WLS}}.\label{eq:power_estimator}
\end{equation}
 By recovering the lost power data, the EMS  can also monitor the unobservable part of the system \cite{Donmez2016Abur}.

\subsection{Optimization of the sampling policy} \label{sec;sampling}
Sensor locations have a significant impact on the estimation performance in power systems  \cite{zhao2014identification}. Therefore,
 in this subsection, we design a sensor allocation policy
 for the  model in  \eqref{reduced model} that aims to minimize the  mean-squared-error (MSE) of the GSP-WLS estimator, where the MSE of an estimator $\hat{\thetavec}$  is 
\beqna
\label{MSE1}
{\text{MSE}}(\hat{\thetavec})={\rm{E}}[(\hat{\thetavec}-\thetavec)^T(\hat{\thetavec}-\thetavec)].
\eeqna
 However, it can be shown that the MSE of $\hat{\thetavec}{}^{\text{GSP-WLS}}$ is a function of the unknown state vector, $\thetavec$, and thus, cannot be used as an objective function for the optimization of the sensor locations. 
 Therefore, we replace the MSE by the 
  Cram$\acute{\text{e}}$r-Rao  bound (CRB) \cite{Kayestimation}, which is a lower bound on the MSE.


In this subsection,  we treat the MSE, bias, and CRB of the vector $\thetavec_{\bar{\mathcal{V}}}$ (without the reference bus) for the sake of simplicity. By substituting  $\theta_1=0$ in the model in \eqref{reduced model} we obtain that the partial measurement vector obtained from a sensor subset $\mathcal{S}$  is a
Gaussian vector with mean $\Hmat_{\mathcal{S},\bar{\mathcal{V}}}\thetavec_{\bar{\mathcal{V}}}$ and  covariance $\Rmat_{\mathcal{S}}$
\beqna\label{pdf}
\zvec_{\mathcal{S}}\sim\mathcal{N}(\Hmat_{\mathcal{S},\bar{\mathcal{V}}}\thetavec_{\bar{\mathcal{V}}},\Rmat_{\mathcal{S}}).
\eeqna
The CRB for this Gaussian vector,
which is a lower bound on the MSE,
is given by
(pp. 45-46 in \cite{Kayestimation}):
\beqna\label{bias_CRB}
{\text{CRB}}({\mathcal{S}})\define
\hspace{5.5cm}\nonumber\\ \text{Tr}\left((\Imat+\frac{\partial \bvec(\mathcal{S})}{\partial\thetavec_{\bar{\mathcal{V}}}})
(\Hmat_{\mathcal{S},\bar{\mathcal{V}}}^T\Rmat_{\mathcal{S}}^{-1}\Hmat_{\mathcal{S},\bar{\mathcal{V}}})^\dagger(\Imat+\frac{\partial \bvec(\mathcal{S})}{\partial\thetavec_{\bar{\mathcal{V}}}})^T\right),
\eeqna
where 
 $\bvec(\mathcal{S})\define {\rm{E}}[\hat{\thetavec}_{\bar{\mathcal{V}}}-\thetavec_{\bar{\mathcal{V}}}]$ is the bias of the estimator  and 
$\frac{\partial \bvec(\mathcal{S})}{\partial\thetavecsmall_{\bar{\mathcal{V}}}}$ is its
 gradient.
Using the model in \eqref{reduced model} and the estimator in \eqref{estimator}, we obtain that
the bias of the GSP-WLS estimator
 is
\beqna\label{bias}
\bvec(\mathcal{S})={\rm{E}}[\hat{\thetavec}_{\bar{\mathcal{V}}}^{\text{GSP-WLS}}-\thetavec_{\bar{\mathcal{V}}}]=\tilde{\Kmat}(\mathcal{S},\mu)\Hmat_{\mathcal{S},\bar{\mathcal{V}}}{\thetavec}_{\bar{\mathcal{V}}}-{\thetavec}_{\bar{\mathcal{V}}},
\eeqna
where 
$\tilde{\Kmat}(\mathcal{S},\mu)$ is defined in \eqref{estimator_matrix}.
Thus, the gradient of \eqref{bias} w.r.t. $\thetavec_{\bar{\mathcal{V}}}$ is
\beqna\label{bias_grad}
\frac{\partial \bvec(\mathcal{S})}{\partial\thetavec_{\bar{\mathcal{V}}}}=\tilde{\Kmat}(\mathcal{S},\mu)\Hmat_{\mathcal{S},\bar{\mathcal{V}}}-\Imat.
\eeqna
By substituting  \eqref{bias_grad} in \eqref{bias_CRB}, we obtain that the CRB on the MSE  of estimators with the GSP-WLS bias  is given by
\beqna\label{CRB2}
{\text{CRB}}({\mathcal{S}})=\hspace{6cm}\nonumber\\
\text{Tr}\Big(\tilde{\Kmat}(\mathcal{S},\mu)\Hmat_{\mathcal{S},\bar{\mathcal{V}}}{(\Hmat_{\mathcal{S},\bar{\mathcal{V}}}^T\Rmat_{\mathcal{S}}^{-1}\Hmat_{\mathcal{S},\bar{\mathcal{V}}})}^{\dagger}\Hmat_{\mathcal{S},\bar{\mathcal{V}}}^T\tilde{\Kmat}^T(\mathcal{S},\mu)\Big).
\eeqna
By substituting \eqref{estimator_matrix} in \eqref{CRB2} and using the pseudo-inverse property, $\Amat=\Amat\Amat^{\dagger}\Amat$, we obtain 
\begin{equation}
    {\text{CRB}}({\mathcal{S}})= \text{Tr}\big(\tilde{\Kmat}(\mathcal{S},\mu)\Rmat_{\mathcal{S}}\tilde{\Kmat}^T(\mathcal{S},\mu)\big).\label{MSE_bound}
\end{equation}

The CRB in \eqref{MSE_bound} 
is not a function of the unknown state vector, $\thetavec$ and
can be used as  an optimization criterion  for choosing the optimal  sensor locations.
We assume a constrained amount of sensing resources, e.g. due to a limited energy and communication budget.
We thus state the problem of the selection of sensor locations with only $\tilde{q}$ sensors as follows:
\beqna
\label{prob_0}
\mathcal{S}^{opt}=\arg\min_{\mathcal{S}\subset\mathcal{M}}{\text{CRB}}({\mathcal{S}})~{\text{s.t. }} |\mathcal{S}|=\tilde{q}\hspace{2.4cm}
\nonumber\\
=\arg\min_{\mathcal{S}\subset\mathcal{M}}\big(\tilde{\Kmat}(\mathcal{S},\mu)\Rmat_{\mathcal{S}}\tilde{\Kmat}^T(\mathcal{S},\mu)\big)~{\text{s.t. }} |\mathcal{S}|=\tilde{q},
\eeqna
where
the last equality is obtained by substituting \eqref{MSE_bound}.

We assume that in the measured buses all the relevant power measurements are given, including active power injections  and flows at the bi-directional connected transmission lines. Thus,  $\mathcal{S}$ is uniquely determined by the chosen buses to measure.
For the sake of simplicity, in the optimization approach, we take ${\Hmat}_{\mathcal{V},\bar{\mathcal{V}}}=\Lmat_{\mathcal{V},\bar{\mathcal{V}}}$ and replace the selection of $\tilde{q}$ sensors by the selection of  $q$ buses.
Thus,  by substituting ${\Hmat}_{\mathcal{V},\bar{\mathcal{V}}}=\Lmat_{\mathcal{V},\bar{\mathcal{V}}}$, we
replace the problem in \eqref{prob_0}  by the  problem of selecting the optimal buses  in the CRB sense.
However, finding the set of $q$ locations among
all the $N$ buses with the smallest CRB  is a combinatorial optimization which has a computational complexity of $\binom{N}{q}$, which is practically infeasible.
Therefore, we propose a  greedy algorithm in Algorithm \ref{Algorithm} for practical implementation of the sampling scheme. The idea behind this algorithm is to iteratively add to the sampling set those buses  that lead to the minimal CRB.
\begin{algorithm}[hbt]
	\algorithmicrequire{\\ 1) Laplacian matrix, $\Lmat$, and noise covariance matrix, $\Rmat$\\
2) Number of buses with sensors, $q$\\ 3)   Regularization parameter, $\mu$}\\
	\algorithmicensure{ Subset of $q$ buses, $\mathcal{S}$}
	\begin{algorithmic}[1]
	\STATE Initialize the bus subset $\mathcal{S}^{(0)}=\emptyset$ and the iteration, $i=0$
		 \WHILE{$i<q$}
		 \STATE Update the set of available locations, $\mathcal{L}=\mathcal{V}\setminus\mathcal{S}^{(i)}$
	 \STATE Find the optimal bus to add:
\beqna\label{objective}
w^{opt}  = \arg \min_{w\in {\mathcal{L}}} \text{Tr}(\{\tilde{\Kmat}(\mathcal{S}^{(i)}\cup w\},\mu)\Rmat_{(\mathcal{S}^{(i)}\cup w)}\\\nonumber\times\tilde{\Kmat}^T(\{\mathcal{S}^{(i)}\cup w\},\mu)),
 \eeqna
where $\tilde{\Kmat}$ is defined in \eqref{estimator_matrix} with ${\Hmat}_{\mathcal{V},\bar{\mathcal{V}}}=\Lmat_{\mathcal{V},\bar{\mathcal{V}}}$ 
 \STATE  Update the subset of buses, $\mathcal{S}^{(i+1)}
\leftarrow \mathcal{S}^{(i)}\cup w^{opt}$, and the iteration, $i\leftarrow i+1$
\ENDWHILE
\STATE 	Update the chosen subset of buses: $\mathcal{S}=\mathcal{S}^{(i)}$
	\end{algorithmic}
	\caption{{Greedy selection of the measured buses}
	\label{Algorithm}}
\end{algorithm}


\section{Extension to the AC-PF model}\label{sec;ac_model}
Since the problem of  low-observability mainly occurs in distribution systems, which requires AC state estimation, in this section we extend the GSP-WLS estimator 
 to the AC-PF model, where we  estimate voltage magnitudes as well.
In particular, we  describe the conventional PSSE in the AC-PF model in Subsection \ref{ACPF_model}.   Then, in Subsection \ref{subsec;state_est_ac}  we present the proposed iterative regularized Gauss-Newton method 
that exploits the smoothness property of the voltage phases and magnitudes in each iteration. 
In Subsection \ref{subsec;remarks_ac} we discuss the properties of the proposed GSP Gauss-Newton algorithm.
\subsection{Model, state estimation, and observability
}
\label{ACPF_model}
In the following, 
we replace the DC-PF model from \eqref{DC_model_vec} 
by the following nonlinear AC-PF model equations:
\begin{equation}\label{eq:ac_model}\zvec=\hvec(\xvec)+\evec,\end{equation}
where
\begin{itemize}[leftmargin=0cm,labelsep=0.1cm,align=left]
    \item $\zvec \in\mathbb{R}^M$ is
the measurement vector that includes the active and reactive branch power
flows and power injections.
\item $\hvec(\xvec)$ is the measurement function, which is determined by the
sensor types and their locations in the network.
\item 
 $\xvec = [\theta_2,\dots,\theta_N,|v_2|,\dots,|v_N|]^T\in\mathbb{R}^{2N-1}$, is the state vector here, where bus $1$ is the reference bus and, thus, $\theta_1=0$ and $v_1$ is known (Chapter 4 \cite{Abur_Gomez_book}).
 \end{itemize}
The specific
forms and parameters of \eqref{eq:ac_model} with different levels of modeling
details can be found e.g. in Chapter 2 of \cite{Abur_Gomez_book}.

Similar to the  WLS estimator  for the DC-PF model in \eqref{WLS}, the  AC-PF state estimation is usually based on  minimizing the following WLS objective function: 
\beqna\label{eq;obj_ac}
   J(\zvec,\hvec(\xvec),\Rmat)=(\zvec-\hvec(\xvec))^T\Rmat^{-1}(\zvec-\hvec(\xvec)),
\eeqna
w.r.t. $\xvec$ \cite{Abur_Gomez_book}.
The first-order optimality condition 
 for the unconstrained
minimization problem in \eqref{eq;obj_ac} is given by
\beqna \label{eq:g_definition}
\gvec(\xvec)\define\frac{\partial J(\zvec,\hvec(\xvec),\Rmat)}{\partial\xvec}=-\Hmat^{T}(\xvec) \Rmat^{-1}(\zvec-\hvec(\xvec))=\zerovec,
\eeqna
where  $\Hmat(\xvec)\define\frac{\partial\hvec(\xvec)}{\partial \xvec}$ is the  Jacobian  matrix of measurement functions $\hvec(\xvec)$ at $\xvec$.
The nonlinear equation, $\gvec(\xvec)=\zerovec$, can be solved using
the (approximated) Gauss-Newton method \cite{Abur_Gomez_book,Monticelli1999}, which results in the following iterative
system: 
\begin{equation}\label{eq:iteration_step}
    \xvec^{(i+1)}=\xvec^{(i)} +\Gmat^{-1}(\xvec^{(i)})\Hmat^{T}(\xvec^{(i)}) \Rmat^{-1}(\zvec-\hvec(\xvec^{(i)})),
\end{equation}
where $\xvec^{(i)}$ is the state estimator at the $i$th iteration
and \beqna
\label{eq:gain1}
\Gmat(\xvec)=\Hmat^T(\xvec)\Rmat^{-1}\Hmat(\xvec)
\eeqna 
is  the gain matrix.
Solving this equation  and iterating until the required accuracy $\varepsilon$ is reached, i.e. until $||\xvec^{(i+1)}-\xvec^{(i)}||\leq\delta$, one will obtain the solution of PSSE.

 The observability requirement for the AC-PF model can be defined by as follows
 (see Chapter 4.6 in \cite{Abur_Gomez_book} 
 and  \cite{Donti2020Zhang}).
 \begin{definition}
 \label{def_obs_ac}
 Assume the AC-PF model from \eqref{eq:ac_model}. 
The network is observable if $\Gmat(\xvec)$  is a nonsingular
matrix for any $\xvec$ in the solution space.
 \end{definition}
 By observing \eqref{eq:gain1}, it can be seen that  if $\Hmat(\xvec)$  has a full column rank of $2N-1$, then the network is observable in the AC-PF sense.
   This observability condition should be satisfied in each iteration of the Gauss-Newton iterative algorithm.
\subsection{GSP-based Gauss-Newton algorithm}\label{subsec;state_est_ac}
Similar to Subsection \ref{sec;state_est}, 
we consider the case where only partial observations of  $\zvec$ from \eqref{eq:ac_model} are available
over  a subset of sensors 
$\mathcal{S}\subseteq {\mathcal{M}}$. 
That is, based on the  model in \eqref{eq:ac_model}, the partial measurement AC-PF model can be written  as
\beqna\label{eq:ac_reduced}
\zvec_{\mathcal{S}}=\hvec_{\mathcal{S}}(\xvec)+\evec_{\mathcal{S}},
\eeqna
where $\evec_{\mathcal{S}}$ 
is a zero-mean Gaussian noise vector
 with a covariance matrix $\Rmat_{\mathcal{S}}$,  as in \eqref{reduced model}.
The
 Jacobian matrix of the model in \eqref{eq:ac_reduced} is   $\Hmat_{\mathcal{S},\bar{\bar{\mathcal{V}}}}(\xvec)=\frac{\partial \hvec_{\mathcal{S}}(\xvec)}{\partial\xvec}$ where $\bar{\bar{\mathcal{V}}}$ indicates the set of all the columns in $\Hmat(\xvec)$.
If the columns of $\Hmat_{\mathcal{S},\bar{\bar{\mathcal{V}}}}(\xvec)$ are linearly dependent, then 
$\Gmat(\xvec)$ is a singular matrix, and from Definition \ref{def_obs_ac}
 the new system
in \eqref{eq:ac_reduced} is {\em{unobservable}}. 
In this case,  the Gauss-Newton iterative procedure for the minimization of  \eqref{eq;obj_ac} cannot be implemented, since  the  update of the solution
cannot be uniquely determined 
from \eqref{eq:iteration_step}.

In order to tackle this problem, we incorporate
 the smoothness constraints from \eqref{eq:smooth_reduce} and \eqref{smothness_full} with $\avec=\vvec$. Thus, the GSP-WLS estimator for the AC-PF model is defined by
\beqna\label{eq:Smooth_wls_ac}
\hat{\xvec}{}^{\text{GSP-WLS}}=\arg\min_{{{\xvec}=[\thetavecsmall^T_{\bar{\mathcal{V}}},\vvec^T]^T\in \mathbb{R}^{2N-1}}}
J(\zvec_{\mathcal{S}},\hvec_{\mathcal{S}}(\xvec),\Rmat_{\mathcal{S}})
\nonumber\\
{\text{such that }} 1)~{\thetavec}_{\bar{\mathcal{V}}}^T {\Lmat}_{\bar{\mathcal{V}}}  {\thetavec}_{\bar{\mathcal{V}}} \leq \varepsilon_ {\thetavecsmall}\hspace{2.8cm}\nonumber\\ {\text{ and }} 2)~({\vvec}_{\bar{\mathcal{V}}}-v_1\onevec)^T {\Lmat}_{\bar{\mathcal{V}}}  ({\vvec}_{\bar{\mathcal{V}}}-v_1\onevec) \leq \varepsilon_{\vvec},
\eeqna
where the function $J$ is defined in \eqref{eq;obj_ac}, and  $\varepsilon_{\thetavecsmall},\varepsilon_{\vvec}$ are the tuning parameters of the smoothness of $\thetavec$ and $\vvec$. 

Then,
by using the KKT conditions \cite{Boyd_2004}, the minimization problem in \eqref{eq:Smooth_wls_ac} can be replaced by the following regularized optimization problem 
:
\beqna\label{eq:min_pro_ac}
\hat{\xvec}{}^{\text{GSP-WLS}}=\hspace{-0.1cm}\arg\min_{{{\xvec}\in \mathbb{R}^{2N-2}}}J_{\text{reg}}(\zvec_{\mathcal{S}},\hvec_{\mathcal{S}}(\xvec),\Rmat_{\mathcal{S}},\bar{\Lmat}),
\eeqna
where \beqna\label{eq;J_reg_definition}
J_{\text{reg}}(\zvec,\hvec(\xvec),\Rmat,\bar{\Lmat})\define J (\zvec,\hvec(\xvec),\Rmat)+{(\xvec-\xvec_0)}^T \bar{\Lmat}{(\xvec-\xvec_0)},
\eeqna
{\small{\begin{equation}\label{eq:def_L_bar}
    \bar{\Lmat}\define\begin{bmatrix}\mu_{\thetavecsmall}\Lmat_{\bar{\nu}} & \zerovec \\ \zerovec & \mu_{\vvec}\Lmat_{\bar{\nu}}
    \end{bmatrix},
\end{equation}}}
and ${\xvec}_{0} \define [\zerovec^T,v_1\onevec^T]^T$.
The term ${\xvec}^T \bar{\Lmat}  {\xvec}$ is a  regularization term, which is based on the smoothness property of the phases and magnitudes of the voltages, established in Section \ref{sub_sec_smooth}.
The parameters $\mu_{\thetavecsmall},\mu_{\vvec}\geq 0$ are Lagrange multipliers that replace $\varepsilon_{\thetavecsmall},\varepsilon_{\vvec}$ as regularization parameters, and their tuning is  discussed in Subsection \ref{subsec;remarks_ac}.

The minimum of the quadratic objective, $J_{\text{reg}}(\zvec_{\mathcal{S}},\hvec_{\mathcal{S}}(\xvec),\Rmat_{\mathcal{S}},\bar{\Lmat})$, can be determined using the first order optimality conditions as follows:
\beqna \label{eq:J_reg_definition}
\gvec_{\text{reg}}(\xvec)\define\frac{\partial J_{\text{reg}}(\zvec_{\mathcal{S}},\hvec_{\mathcal{S}}(\xvec),\Rmat_{\mathcal{S}},\bar{\Lmat})}{\partial\xvec}\hspace{3.35cm}
\nonumber\\=-\Hmat_{\mathcal{S},\bar{\bar{\mathcal{V}}}}^{T}(\xvec) \Rmat_{\mathcal{S}}^{-1}(\zvec_{\mathcal{S}}-\hvec_{\mathcal{S}}(\xvec))+\bar{\Lmat}(\xvec-\xvec_0)=\zerovec.
\eeqna
Then, similar to \eqref{eq:iteration_step},
the nonlinear equation, $\gvec_{\text{reg}}(\xvec)=\zerovec$, is  solved using
the (approximated) Gauss-Newton method,  which results in this case in the following iterative
system:
\beqna\label{eq:iteration_step_regulrized}
   \xvec^{(i+1)}=\xvec^{(i)}+
    \Gmat_{\text{reg}}^{-1}(\xvec^{(i)})\hspace{4cm}\nonumber\\\times\left(\Hmat_{\bar{\bar{\mathcal{V}}},\mathcal{S}}^{T}(\xvec^{(i)}) \Rmat_{\mathcal{S}}^{-1}(\zvec_{\mathcal{S}}-\hvec_{\mathcal{S}}(\xvec^{(i)}))
    -\bar{\Lmat}(\xvec^{(i)}-\xvec_0)\right),
\eeqna
where \beqna\label{eq;regularized_G}
\Gmat_{\text{reg}}(\xvec)\define\frac{\partial \gvec_{\text{reg}}(\xvec)}{\partial\xvec}=\Hmat_{\mathcal{S},\bar{\bar{\mathcal{V}}}}^{T}(\xvec) \Rmat_{\mathcal{S}}^{-1}\Hmat_{\mathcal{S},\bar{\bar{\mathcal{V}}}}(\xvec)+\bar{\Lmat}
\eeqna is the new gain matrix.
Solving this equation  and iterating until the required accuracy $\delta$ is reached, i.e. $||\xvec^{(i+1)}-\xvec^{(i)}||\leq\delta$, one will obtain the  proposed GSP-WLS estimator for the AC-PF model.
It can be seen that for an unobservable system, the matrix $\Hmat_{\mathcal{S},\bar{\bar{\mathcal{V}}}}^{T}(\xvec) \Rmat_{\mathcal{S}}^{-1}\Hmat_{\mathcal{S},\bar{\bar{\mathcal{V}}}}(\xvec)$ is a singular matrix and the additional  terms in \eqref{eq;regularized_G}, $\bar{\Lmat}$ from \eqref{eq:def_L_bar} with $\mu_{\thetavecsmall}> 0$, and/or $\mu_{\vvec}> 0$, can enable the matrix inversion of $\Gmat_{\text{reg}}(\xvec)$ and improve the numerical stability of the GSP-WLS AC-PF model based estimator.
The iterative solution is summarized in Algorithm \ref{Algorithm2}. For the initialization, we suggest to use the ``flat voltage profile", in which all bus voltages are assumed to be $1.0$ per unit and in phase with each other \cite{Abur_Gomez_book}.
\begin{algorithm}[hbt]
	\textbf{Input:}{\\ 1) Laplacian matrix, $\Lmat$, and noise covariance matrix, $\Rmat_{\mathcal{S}}$\\
2) Tuning  parameters: $\delta$,  $\mu_{\thetavecsmall}$, $\mu_{\vvec}$} and number of iterations, $l$\\ 3)    Measurements vector, $\zvec_{\mathcal{S}}$, and the function, $\hvec_{\mathcal{S}}(\cdot)$
	\newline\textbf{Output:}{State estimator, $\hat{\xvec}$}
	\begin{algorithmic}[1]
	\STATE Initialize the state vector $\xvec^{(0)}$
			 \FOR{$i=0,\dots,l$}
 \STATE Calculate the right hand side of \eqref{eq:iteration_step_regulrized} for $\xvec^{(i)}$
 \STATE  Solve \eqref{eq:iteration_step_regulrized} for  $\xvec^{(i+1)}$
 \STATE \textbf{if} $||\xvec^{(i+1)}-\xvec^{(i)}||<\delta$: \textbf{break}
	\ENDFOR
	\STATE 	Update the state vector: $\hat{\xvec}=\xvec^{(i+1)}$
	\end{algorithmic}
	\caption{{Regularized Gauss-Newton (GSP-WLS)} \label{Algorithm2}}
\end{algorithm}

\subsection{Discussion}
\label{subsec;remarks_ac}
In the following, we present special cases of the proposed GSP-WLS estimator for the AC-PF model implemented by the regularized Gauss-Newton method.
\begin{enumerate}[leftmargin=0.5cm,labelsep=0.25cm,align=left]
\item\label{special_case_1_ac} \underline{Full observability:} If we have access to all sensors, i.e. if  $\hvec_{\mathcal{S}}(\xvec)=\hvec(\xvec)$, then, by substituting 
 $\mu_{\thetavecsmall}=\mu_{\vvec}=0$ in
\eqref{eq:iteration_step_regulrized} we obtain that
$\Gmat_{\text{reg}}(\xvec)= \Gmat(\xvec)$,
where $\Gmat(\xvec)$ and 
$\Gmat_{\text{reg}}(\xvec)$
are defined in \eqref{eq:gain1} and \eqref{eq;regularized_G}.
Thus, in this case,  the Gauss-Newton iteration of the GSP-WLS estimator in \eqref{eq:iteration_step_regulrized} coincides with the Gauss-Newton iteration in \eqref{eq:iteration_step}.

\item \underline{Relation with the pm-WLS estimator:}
Similar to the DC-PF model, the pm-WLS estimator for unobservable systems is calculated based on  the  measurements and the {\em{a-priori}} estimated (predicted) states,  $\hat{\xvec}_{prior}\in\mathbb{R}^{ 2N-1}$, with the forecasting error covariance matrix, $\Rmat_{prior}$.
 Thus, the following pm-WLS estimator is used  \cite{5233865}:
\beqna\label{eq:min_pro_pseudo_ac}
\hat{\xvec}{}^{\text{pm-WLS}}=\arg\min_{{{\xvec}\in \mathbb{R}^{2N-1}}} \left.\{J (\zvec_{\mathcal{S}},\hvec_{\mathcal{S}}(\xvec),\Rmat_{\mathcal{S}})
\right.\hspace{1.5cm}
\nonumber\\
\left.+({\xvec}-\hat{\xvec}_{prior})^T \Rmat_{prior}^{-1}({\xvec}-\hat{\xvec}_{prior})\right\}.
\eeqna
The pm-WLS estimator for the AC-PF model can be calculated with the Gauss-Newton method \cite{5233865}.
It can be seen that if we substitute $\hat{\xvec}_{prior}=\xvec_0$  and $\Rmat_{prior}^{-1}=\bar{\Lmat}$ 
then the pm-WLS estimator from \eqref{eq:min_pro_pseudo_ac} coincides with the GSP-WLS estimator from \eqref{eq:min_pro_ac}-\eqref{eq;J_reg_definition}. 
Thus, similar to the DC-PF model, the proposed GSP-WLS estimator can be interpreted as a special case of the pm-WLS estimator, where the GSP theory gives the mathematical justification for the determination of the pseudo-data information.
\end{enumerate}

\section{Simulations}
\label{sec;Simulation}
In this section, the performance of the  proposed methods is investigated.
In Subsection \ref{state_est_subsec}, we evaluate the performance of the  GSP-WLS estimator from Section \ref{sec;state_est}. In Subsection \ref{ac_state_est_subsec}, we evaluate the performance of the regularized Gauss-Newton method GSP-WLS (AC-model based) from Section \ref{sec;ac_model}.
The influence of 
 the sampling policy from Subsection \ref{sec;sampling} is examined for both cases.

In all simulations, the measurements were generated according to the AC model  \cite{Abur_Gomez_book}, with the parameters taken from   \cite{iEEEdata} for the IEEE 118-bus system, which has $N=118$ 
buses and at most $M=952$ measurements (of active and reactive power). The state estimator of  the reference bus is set to $\hat{\theta}_1=0$, and the  noise covariance matrix to $\Rmat=\sigma^2\Imat$, where, unless
otherwise stated, $\sigma^2=0.01$. The regularization parameters are $\mu=0.1$ for the estimator in \eqref{estimator}-\eqref{estimator_matrix}, and $\mu_{\theta}=0.045$, $\mu_{\vvec}=10$ in \eqref{eq:min_pro_ac}-\eqref{eq:def_L_bar}.
The performance is evaluated using  $1,000$ Monte-Carlo simulations, unless otherwise stated.

\begin{figure}[hbt]
		\begin{center}
		\centering\includegraphics[width=5cm]{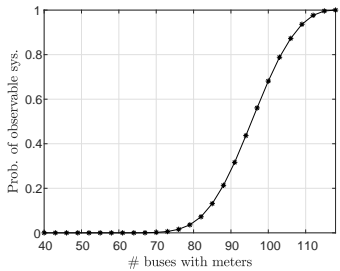}
		\caption{The estimated probability of  the IEEE 118-bus system to be observable versus the number of measured buses.}
		\label{fig:probability_2_be_observable}	
		\end{center}
	\end{figure}

In order to demonstrate the system observability,
in Fig. \ref{fig:probability_2_be_observable} we present the estimated probability of the system to be observable, according to Definition \ref{def_obs}, versus the number of  measured buses.
 The estimated probability of observability is calculated as the  percentage of  the scenarios of observable systems in $100,000$ Monte-Carlo simulations for randomly selected buses in the system.
It can be seen that this probability  increases as the number of measured buses increases and that the IEEE 118-bus system becomes unobservable with probability $1$ for less than $72$ measured buses.
\subsection{State estimation and sampling  under the DC-PF model}
\label{state_est_subsec}
In the following, we evaluate the performance of the GSP-WLS estimator from \eqref{estimator} and  compare it with the performance of the pm-WLS estimator from \eqref{eq:pseudo_estimator}.
The pm-WLS estimator was generated as follows: First, the {\em{a-priori}} estimated  states, $\hat{\thetavec}_{prior}$, were generated from a zero-mean Gaussian distribution with covariance  $0.015\Imat$.
Second,
for each simulation 
we implement the estimator in \eqref{eq:pseudo_estimator}-\eqref{K_2} by using the random value of $\hat{\thetavec}_{prior}$
and the covariance $\Rmat_{prior}^{-1}=0.5\Imat$.

We compare the estimation performance 
for the following bus selection policies:
\begin{enumerate}[leftmargin=0.2cm,labelsep=0.2cm,align=left,label={{\bf{\roman*)}}~}]
    \item Random bus selection policy (rand.) -  the measured buses  are randomly chosen 
 independently  from $\{1,\ldots,N\}$,
 where for more than $72$ buses only observable systems are taken.
\item Experimentally designed sampling (E-design) \cite{chen2015discrete} -  the basic assumption behind this method is that the measured graph signal (here the power signal) is an $R$ bandlimited  signal in the graph frequency domain. That is,  the GFT of $\zvec_{\text{bus}}$, as defined in \eqref{GFT}, satisfies
$(\tilde{z}_{\text{bus}})_n=0$,  $n=R+1,\ldots,N$, where $R$ is the cutoff frequency.  Practically, this method maximizes the smallest singular value of the matrix $\Vmat_{{\mathcal{S}},\{1,\ldots,R\}}$, where  we set  $R=48$, by trial and error.
Since in practice the $R$-low-pass assumption does not hold for the power injection signal  (as can be seen in Fig. \ref{fig31}), this  commonly-used method that was suggested 
 in \cite{bandlimite_PSSE} for power systems, is expected to yield poor performance.
    \item  Minimum  CRB (Alg. 1) - the proposed bus selection policy from Algorithm \ref{Algorithm}, which does not necessarily result in an  observable system.
\end{enumerate}

\begin{figure}[htb]
    \centering
    \subcaptionbox{\label{fig:MSE_vs_setSize}}[\linewidth]
    {\includegraphics[width=7cm]{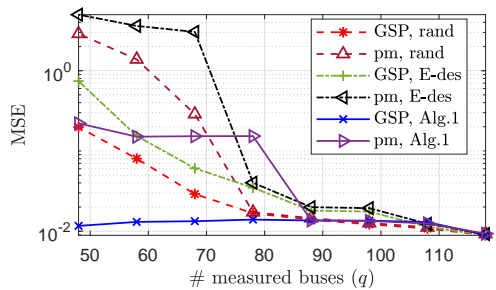}}
    \subcaptionbox{\label{fig:MSE_vs_SNR}}[\linewidth]
    {\includegraphics[width=6.5cm]{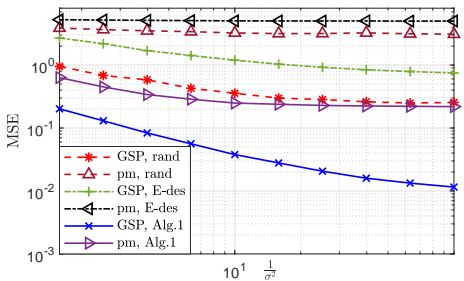}}
    \caption{
    State estimation under the DC-PF model: The MSE of the GSP-WLS  and the pm-WLS estimators for 
    random, E-design, and Algorithm \ref{Algorithm} bus selection methods, for the IEEE 118-bus system  versus: (a) the number of  buses, $q$, with $\sigma^2=0.01$; and (b) $\frac{1}{\sigma^2}$ with $q=48$  buses.}
     \label{fig:PSSE}
\end{figure}

In Figs. \ref{fig:MSE_vs_setSize} and  \ref{fig:MSE_vs_SNR} the MSE of the GSP-WLS estimator  and  of the pm-WLS estimator  are presented 
 versus the number of measured buses, $q$, and  versus $\frac{1}{\sigma^2}$, respectively, 
 with the sampling policies  1)-3). 
Figure \ref{fig:MSE_vs_SNR} is obtained for
 $q=48$ measured buses, for which the system is unobservable with  probability almost 1.
 It can be seen that
 the MSE decreases as the number of measured buses increases and as $\sigma^2$ decreases.
 In both figures,
the GSP-WLS estimator outperforms the pm-WLS estimator for any tested sampling policy.
 In Fig. \ref{fig:MSE_vs_setSize}, it can be seen that
for each sampling policy, the MSE of the GSP-WLS  and the pm-WLS estimators separate from each other where the system becomes  unobservable  (i.e. where $q<72$ for the random sampling, $q<76$ for the E-design sampling, and $q<79$ for Algorithm \ref{Algorithm}). 
In addition, it can be seen that the proposed sampling policy from Algorithm \ref{Algorithm} results in a significantly lower MSE than that obtained for the random  and the E-design sampling policies for both  estimators.
In Fig.   \ref{fig:MSE_vs_SNR} we can see that
for small values of $\sigma^2$, the differences in the MSE performance of the two estimators are than those obtained for large values of $\sigma^2$. 
This is because the noise impairs the smoothness of the state signal.

\begin{figure}[htb]
		\begin{center}
		\centering\includegraphics[width=7cm]{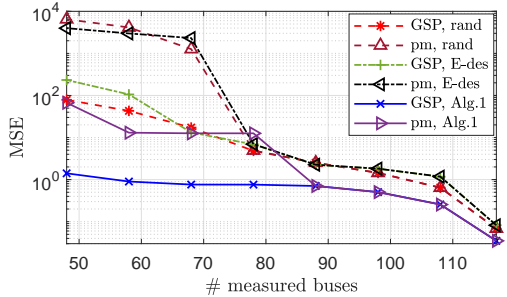}
		\caption{Power estimation based on the DC-PF model: 
		The MSE of the GSP-WLS  and the pm-WLS estimators for  random, E-design, and Algorithm \ref{Algorithm} bus selection methods, for the IEEE 118-bus system and $\sigma^2=0.01$ versus the number of  buses, $q$.}
		\label{fig:mse_power}	
		\end{center}
	\end{figure}

 Figure \ref{fig:mse_power} shows the MSE of the power estimator $\hat{\zvec}$  from \eqref{eq:power_estimator}    with the three sampling policies. 
		 It can be seen  that the MSE decreases as the network size increases, as expected, since  we have fewer  parameters to  estimate with the increase in the number of samples (the estimation error in the measured buses is zero) and the state estimation is more accurate, as we presented in Fig. \ref{fig:MSE_vs_setSize}.
    In addition, the relations between the sampling policies and the estimation methods are similar to in Fig. \ref{fig:MSE_vs_setSize}, where the GSP-WLS estimator with the proposed bus selection policy of Algorithm \ref{Algorithm} achieves the lowest MSE.

\subsection{State estimation and sampling  under the AC-PF model}\label{ac_state_est_subsec}
In the following, we evaluate the performance of the 
regularized Gauss-Newton  method for implementing the
GSP-WLS estimator from Algorithm \ref{Algorithm2}, and  of the pm-WLS estimator from \eqref{eq:min_pro_pseudo_ac} in the estimation of both the phases and the magnitude.
The  pm-WLS estimator was also implemented by  Algorithm \ref{Algorithm2} with the appropriate  replacements of the regularization terms, i.e. where we use $\hat{\xvec}_{prior}$  and $\Rmat_{prior}^{-1}=\Imat$  instead of $\xvec_0$ and  $\bar{\Lmat}$ (see the discussion after \eqref{eq:min_pro_pseudo_ac}). The maximal number of iterations is set to $l=20$ and $\delta=10^{-8}$.

\begin{figure}[htb]
    \centering
    {\includegraphics[width=8cm]{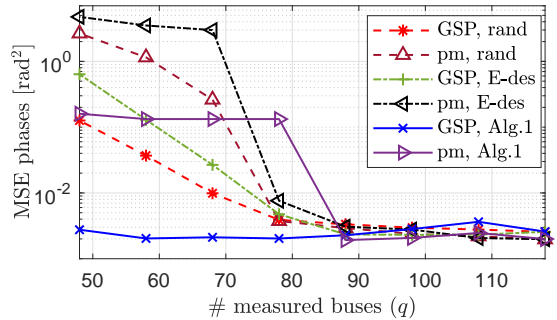}}
    \caption{
    State estimation under the AC-PF model: The phase estimation MSE of the GSP-WLS  and the pm-WLS estimators for 
    random, E-design, and Algorithm \ref{Algorithm} bus selection methods, for the IEEE 118-bus system  versus the number of  buses, $q$.
    }
     \label{fig:PSSE_phases}
\end{figure}
\begin{figure}[htb]
    \centering
    {\includegraphics[width=8cm]{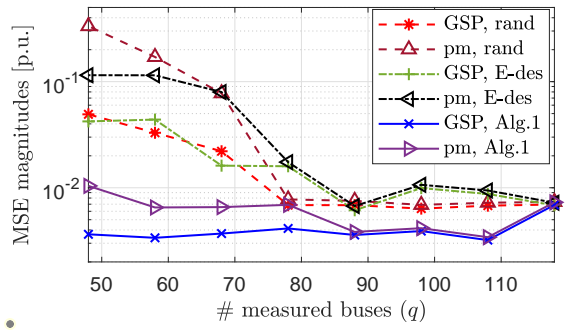}}
    \caption{
    State estimation under the AC-PF model: The magnitude MSE of the GSP-WLS  and the pm-WLS estimators for 
    random, E-design, and Algorithm \ref{Algorithm} bus selection methods, for the IEEE 118-bus system  versus the number of  buses, $q$.
    }
     \label{fig:PSSE_manitudes}
\end{figure}

In Fig. \ref{fig:PSSE_phases}
the MSE of phase estimation by the GSP-WLS estimator (implemented by the regularized Gauss-Newton  method)  and  by the pm-WLS estimator  are presented 
 versus the number of measured buses, $q$,
 for  the sampling policies  1)-3). 
 Similarly, in Fig. \ref{fig:PSSE_manitudes} 
 the MSE of the magnitude estimation is presented.
 It can be seen that
 the MSE  decreases as the number of measured buses increases for both the magnitudes and the phases.
It can be seen that the GSP-WLS estimator outperforms the pm-WLS estimator for any sampling policy.
In Figs. \ref{fig:PSSE_phases} and \ref{fig:PSSE_manitudes}, it can be seen that
 for each sampling policy, the MSE of the GSP-WLS  and the pm-WLS estimators separate from each other where the system becomes  unobservable  (i.e. where $q<72$ for the random sampling, $q<76$ for the E-design sampling, and $q<79$ for Algorithm \ref{Algorithm}). 
Finally, it can be seen that the  sampling policy from Algorithm \ref{Algorithm} results in a significantly lower MSE than that obtained for the random  and the E-design sampling policies for the GSP-WLS estimator.

\section{conclusion}
\label{conclusion}
In this paper, we propose a GSP framework for  state estimation and sensor placement strategy in unobservable systems. 
In particular,
we first validate the graph smoothness of the phases and the magnitudes of the voltages w.r.t. the admittance matrix. Then,
we derive  the GSP-WLS estimator for state estimation in unobservable systems  under both the DC-PF and AC-PF models. 
The GSP-WLS estimator uses the graph-smoothness of the state signals as a regularization term. 
In addition, we introduce a greedy algorithm to tackle the problem of selecting the sampling set that optimizes the state estimation performance.
 Simulation results demonstrate the potential
of the GSP methods in power systems  for cases that are otherwise unobservable. It is shown that the proposed methods  can accurately estimate voltage phasors (or, equivalently, phases and magnitudes)
under low-observability conditions where standard
 methods cannot.
Possible extensions of the proposed GSP framework for power system    include the development of  approaches
based on PMU measurements, the consideration of time-series measurements with temporal dependencies, and the extension to graph neural networks. 
\bibliographystyle{IEEEtran}

\end{document}

%% file: myshorts.tex
\newcommand{\avec}{{\bf{a}}}

\newcommand{\bvec}{{\bf{b}}}

\newcommand{\evec}{{\bf{e}}}

\newcommand{\xvec}{{\bf{x}}}
\newcommand{\zvec}{{\bf{z}}}

\newcommand{\vvec}{{\bf{v}}}
\newcommand{\gvec}{{\bf{g}}}

\newcommand{\hvec}{{\bf{h}}}

\newcommand{\onevec}{{\bf{1}}}
\newcommand{\zerovec}{{\bf{0}}}

\newcommand{\Lambdamat}{{\bf{\Lambda}}}

\newcommand{\Amat}{{\bf{A}}}
\newcommand{\Bmat}{{\bf{B}}}

\newcommand{\Gmat}{{\bf{G}}}
\newcommand{\Hmat}{{\bf{H}}}

\newcommand{\Imat}{{\bf{I}}}
\newcommand{\Lmat}{{\bf{L}}}
\newcommand{\Kmat}{{\bf{K}}}

\newcommand{\Rmat}{{\bf{R}}}

\newcommand{\Vmat}{{\bf{V}}}
\newcommand{\Wmat}{{\bf{W}}}

\newcommand{\define}{\stackrel{\triangle}{=}}

\def\thetavec{{\mbox{\boldmath $\theta$}}}

\def\thetavecsmall{{\mbox{\boldmath {\scriptsize $\theta$}}}}

\newcommand{\be}{\begin{equation}}
\newcommand{\ee}{\end{equation}}
\newcommand{\beqna}{\begin{eqnarray}}
\newcommand{\eeqna}{\end{eqnarray}}
